\documentclass[prl,aps,twocolumn,groupedaddress]{revtex4}
\usepackage{epsfig,latexsym,amssymb,amsmath,amsbsy,graphics,graphicx}

\def \beq{\begin{equation}}
\def \eeq{\end{equation}}
\def \beqarr{\begin{eqnarray}}
\def \eeqarr{\end{eqnarray}}

\begin{document}

\title{Superfluid-Insulator Transition and Fermion Pairing in Bose-Fermi Mixtures
}

\author{Kun Yang}
\affiliation{NHMFL and Department of Physics, Florida State
University, Tallahassee, Florida 32306, USA}

\affiliation{Zhejiang Institute of Modern Physics, Zhejiang
University, Hangzhou 310027, P.R. China}

\date{\today}

\begin{abstract}

It is well known that bosons on an optical lattice undergo a
second-order superfluid-insulator transition (SIT) when the lattice
potential increases. In this paper we study SIT when fermions
coexist with the bosons. We find that the critical properties of
particle-hole symmetric SIT with dynamical exponent $z=1$ is
modified when fermions are present; it either becomes a
fluctuation-driven first order transition or a different
second-order transition. On the other hand the more generic
particle-hole asymmetric (with $z=2$) SIT is stable against coupling
with fermions. We also discuss pairing interaction between fermions
mediated by quantum critical fluctuations near SIT, and show that the fermion
pairing/superfluid transition temperature peaks near SIT when a single species
of fermion is present.

\end{abstract}

\maketitle

\section{introduction}

The observation of superfluid-insulator transition (SIT) of bosons
on optical lattices\cite{greiner,bloch} is arguably one of the most
important recent developments in the field of cold atom physics; it
triggered a tremendous amount of interest and activities in studying
strong correlation physics using optical lattices. More recently,
much work has focused on strong interaction/correlation physics that
involve fermions, like superfluidity resulting from fermion pairing,
and realization of various exotic phases proposed in the context of
electronic condensed matter physics\cite{bloch}. Another direction
of thrust is Bose-Fermi mixtures on optical lattices; these are new
many-body systems that have no direct analogy in electronic
condensed matter systems.

In this paper we consider boson SIT on optical lattices in
Bose-Fermi mixtures in two- and three-dimensions, and consider two
separate situations: (i) Bosonic atoms coexist with fermionic atoms.
(ii) The bosons are molecules made of two fermionic atoms of
different species (or opposite ``spins"), and the numbers of
fermions are different for different spins. Due to the imbalance
some of the fermions are unpaired and coexist with bosonic
molecules. While the physical settings are very different, we show
that the effective field theory describing the SIT turn out to be
the same for both cases. We find that the presence of fermion do {\em not} affect the critical properties of
SIT for most parts of the phase boundary separating the superfluid and insulating phases, provided the Bose-Fermi interaction is sufficiently weak. On the other hand near some special point with (emerging) particle-hole symmetry, fermions do affect the critical properties of SIT. These are the subjects of section II below.

Bosons mediate an attractive interaction between fermions, which leads to fermion pairing and superfluidity at low temperature. In section III we discuss this pairing interaction near boson SIT, which is dominated by critical fluctuations. We show that under certain conditions the fermion superfluid transition temperature peaks near SIT.

A few concluding remarks are offered in section IV.

\section{boson superfluid-insulator transition}

Theoretically, the SIT is well understood in purely bosonic systems
based on simple models like the Bose-Hubbard
model\cite{fisher89,sachdevbook}:
\begin{equation}
H_{B}=-t_b\sum_{<ij>}b^\dagger_{i}b_{j}-\mu_b\sum_i n_{bi} +{U\over
2}\sum_i n_{bi}(n_{bi}-1), \label{bosehubbard}
\end{equation}
where $b_i$ is the boson annihilation operator on site $i$, $n_{bi}=b^\dagger_{i}b_{i}$,
$U > 0$ is the on-site repulsion between the bosons, $t_b$ is the boson hopping matrix
element and $\mu_b$ is boson chemical potential. The
schematic phase diagram of (\ref{bosehubbard}) at zero temperature
($T=0$) can be found in Fig. 1a, where there are two phases: At
small $t_b/U$ the system is a Mott insulator with an integer number
of bosons per site; for larger $t_b/U$ bosons declocalize and Bose
condense, resulting in a superfluid phase. The superfluid-insulator
transition (SIT) is continuous, and depending on the value of boson
chemical potential $\mu_b$ there are two universality classes for
the SIT\cite{sachdevbook}: (a) Generically, either particles or
holes condense at the SIT, and the transition is the dilute Bose gas
transition with dynamical exponent $z=2$; (b) at the tips of the
lobes of the Mott insulator phase, particles and holes condense
simultaneously, and the SIT is in the universality class of $d+1$
dimensional $XY$ transition with $z=1$ (here $d$ is the spatial
dimension of the system). The dynamical exponent $z$ is the critical exponent
that characterizes the
relation between energy/freqeuncy and momentum/wavevector of quantum critical
excitations/fluctuations: $\omega_{k}\sim k^z$. For the two universality classes
of SIT here the values of $z$ can be understood from the following considerations:
(a) For the particle-hole asymmetric case, the critical excitations are either particles
{\em or} holes that become gapless and start to condense at SIT; they have
{\em non-relativistic} dispersion relation $\omega_{k} = k^2/(2m^*)$ ($m^*$ is their
effective mass), and $z=2$ simply reflects this non-relativistic dispersion. (b) In the
particle-hole symmetric case, both particles {\em and} holes become gapless at SIT, and must
be treated on equal-footing. Thus the effective field theory contains both particle and
antiparticle (hole), which turn out to be a {\em relativistic} quantum field theory with
emergent Lorentz symmetry. The Lorentz symmetry requires that space and time are symmetric,
which guarantees that frequency must scale the same way as wavevector: $\omega_{k}\sim k$;
this implies $z=1$.

\begin{figure}
\includegraphics[width=0.45\textwidth]{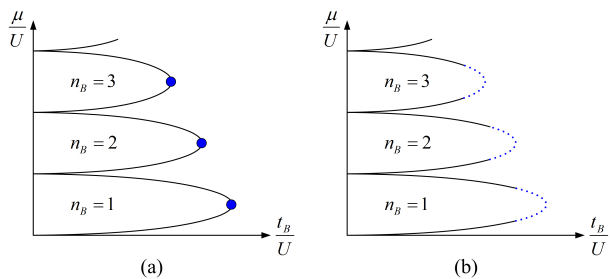}
\caption{(Color on-line) (a) Schematic phase diagram of Bose Hubbard
model. Solid lines are 2nd order phase boundaries separating superfluid phase
from Mott insulator phases with fixed boson number $n_B$ per site. For each
lobe of the Mott insulator characterized by a fixed $n_B$,
the superfluid-insulator transition (SIT) is triggered by the condensation
of either particles (upper half of the lobe) or holes (lower half of the lobe),
except at the tip of the lobe (blue dots) where
particles and holes condense simultaneously. (b) In the presence of
fermions, SIT may become 1st order near the tips of MI lobes (blue
dotted lines).}
\end{figure}

The presence of fermions introduces {\em additional} gapless fermionic
excitations, which can potentially change the critical
properties of these two universality classes.
We find that in the presence of fermions, the coupling between
bosons and gapless fermionic excitations
near the Fermi surface
leads to singular terms in the effective field theory for SIT. Such
singular terms turn out to be irrelevant at the $z=2$ fixed point,
thus this fixed point is stable. On the other hand they lead to
run-away renormalization group (RG) flows at the $z=1$ fixed point,
thus destabilizing this fixed point; as a result near the tips of
the Mott insulator lobes SIT either becomes a fluctuation-driven
first order transition, or becomes a different 2nd order transition.
In next section we will also study the pairing interaction between fermions mediated by
quantum critical fluctuations near SIT, and show that under certain
conditions the fermion pairing gap and transition temperature peaks
near SIT.

We begin our discussion with case (i) of Bose-Fermi mixture,
described by the following Hamiltonian:
\begin{equation}
H=H_{B}-t_f\sum_{<ij>}(f^\dagger_{i}f_{j}+ h.c.) -\mu_f\sum_i n_{fi}
+V\sum_i n_{bi}n_{fi}, \label{bosefermi1}
\end{equation}
where $f_i$ is the fermion operator on site $i$,
$n_{fi}=f^\dag_if_i$, and $V$ is the Bose-Fermi interaction
strength. Even though the Hamiltonian (\ref{bosefermi1}) is still
quite simple, it actually has a very rich phase diagram\cite{lewenstein}.
To avoid complications associated with instabilities like phase separation or bound
state formation, we assume $V$ to be weak and fermion density to be
low. To derive the effective field theory for boson SIT, we use a
functional integral representation of Eq. (\ref{bosefermi1}), and
integrate out the (quadratic) fermionic degrees of freedom to obtain
an effective action in terms of the bosons only. This procedure
generates additional interaction terms for the bosons, in the form
of
\begin{equation}
\sum_n\Lambda_{2n}(q_1,\cdots,q_n;q'_1,\cdots,q'_n)\overline{b}(q_1)\cdots\overline{b}(q_n)b(q'_1)\cdots
b(q'_n). \label{intfermi}
\end{equation}
Here $q=({\bf q},\omega)$ includes both momentum and frequency, and $b$ is the bosonic
field of the effective field theory.
These terms are represented diagramatically in Fig. 2.

\begin{figure}
\begin{tabular}{cccccc}
\parbox[c]{1.7cm}{\includegraphics[scale=0.25]{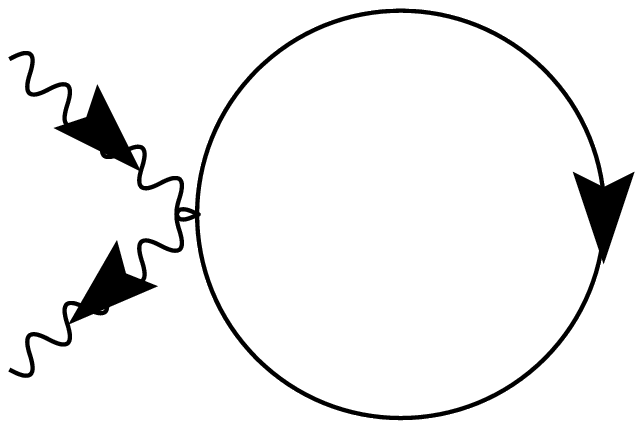}} & + &
\parbox[c]{2.1cm}{\includegraphics[scale=0.25]{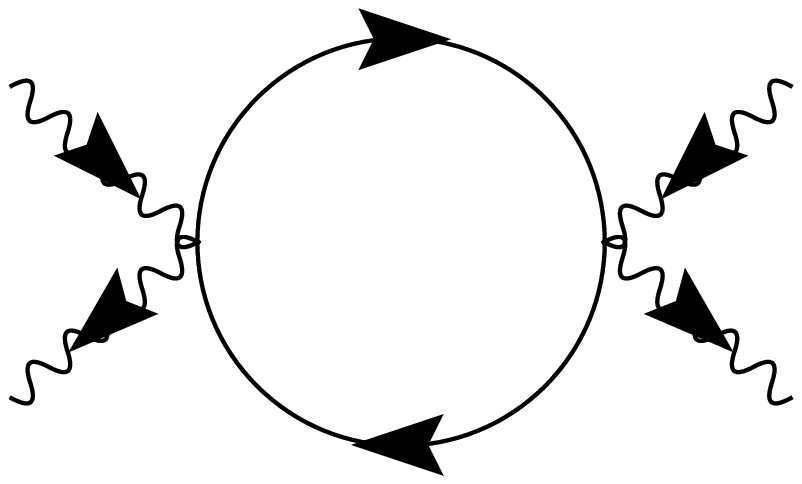}} & + &
\parbox[c]{2.1cm}{\includegraphics[scale=0.25]{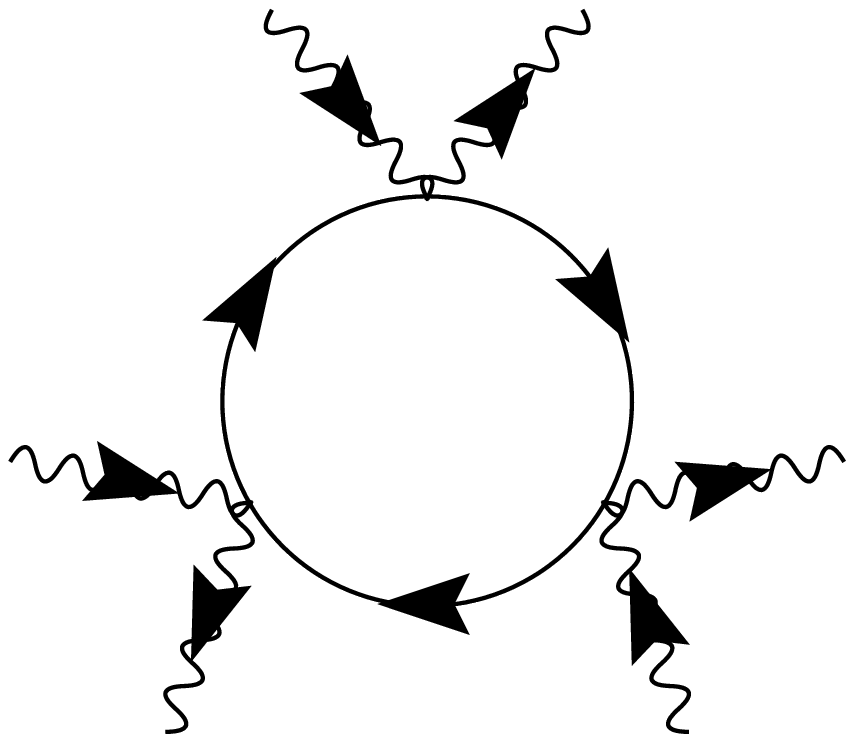}} & + $\cdots$
\end{tabular}
\caption{Feynman diagrams representing terms in effective bosonic
action generated from integrating out fermions, for case (i) of
Bose-Fermi mixture. The solid and wavy lines are fermion and boson
propagators respectively.}
\end{figure}

Because there exist gapless excitations around the Fermi surface,
tracing out fermions leads to terms in ({\ref{intfermi}) with
singular coupling functions $\Lambda_{2n}$ at long wave-length
and/or low frequency (i.e., $\Lambda_{2n}(q)$ is non-analytic at
small ${\bf q}$ and/or $\omega$). The most studied cases with such singularities
generated by tracing out gapless fermions are Hertz
theories for magnetic transitions in metals\cite{hertz}, where due
to the direct coupling between order parameter and fermions,
singularities appear at the quadratic level in the form of Landau
damping. Such singularity not only changes the critical properties
of the phase transition at Gaussian (or tree) level\cite{hertz}, but
also casts the validity of integrating out fermions in doubt\cite{hertznote}.

In the present case however, we find the singularities are less severe.
This is because here the order parameter $b$ does not couple to
fermions directly; instead the coupling is in the form of
density-density interaction between bosons and fermions. As a result
{\em no singularity} appears in the quadratic action (i.e., no
singularity in $\Lambda_{2}(q)$); the bosons only receive a Hartree
shift of chemical potential here. Consequently power counting and
critical behavior do not change at this level.

Singularities do appear in higher order terms of ({\ref{intfermi}),
and the leading singularity is in a quartic term of the form:
\begin{equation}
\lambda_4\int{d{\bf q}d\omega}{|\omega|\over |{\bf q}|}||b|^2({\bf
q},\omega)|^2, \label{singularterm}
\end{equation}
where $|b|^2({\bf q},\omega)$ is the Fourier transform of $|b({\bf
r}, \tau)|^2$.\cite{singularitynote} The importance of this singular quartic term depends
on the dynamical exponent $z$. For $z=2$, $\omega$ scales as $|{\bf
q}|^2$; it is clear that it is much less important than the regular
quartic term $\int{d{\bf r}d\tau}|b({\bf r},\tau)|^4$, and in fact
scales to zero from a simple power counting analysis which leads to
the RG flow for $\lambda_4$:
\begin{equation}
{d \lambda_4\over d\log s}=(5-d-2z)\lambda_4,
\label{rg}
\end{equation}
where $s$ is the length rescaling factor that relates the length before and
after an RG transformation: $x'=x/s$ (the corresponding transformation for
time is $t'=t/s^z$, which provides the technical definition of dynamical
exponent $z$). Clearly $\lambda_4$ is an
irrelevant coupling for $z=2$ and $d > 1$. We thus find that the
$z=2$ quantum critical point (QCP) of the SIT is {\em stable}
against weak Bose-Fermi coupling in this case.

The situation is very different at the particle-hole symmetric QCP
of SIT with $z=1$. In this case Eq. (\ref{rg}) suggests $\lambda_4$
is marginal for $d=3$ while relevant for $d=2$; it can thus change
the critical properties of the SIT at this particle-hole symmetric
QCP. Sachdev and Morinari\cite{sm} analyzed a very similar model
with the same type of singular term in a different context. As they
showed, the $\lambda_4$ coupling is actually non-renormalizable, as
at each step of RG, new relevant or marginal couplings of different
types of singularities appear, as a result of which the standard RG
procedure is unstable. As an alternative they introduced a new field
to decouple the singular coupling (\ref{singularterm}), and
corresponding coupling constants; the theory becomes renormalizable
after such a modification. Within a perturbative RG treatment they
find that the field theory yields no fixed point at weak coupling.

There are two possible interpretations of the absence of fixed point
within perturbative RG for the particle-hole symmetric case. It
could indicate that the SIT is driven first order by fluctuations
due to Bose-Fermi coupling around the tips of the Mott insulator
lobes, as indicated in Fig. 1b. Another more interesting possibility
is that there actually exists a strong-coupling fixed point that is
inaccessible to perturbative RG. If this turns out to be the case
the SIT would remain 2nd order, but the critical properties would be
controlled by the new fixed point at which the Bose-Fermi coupling
strength flows to a universal value; this would dramatically affect
the fermion properties at the transition (to be discussed later). In
this case the phase diagram remains that of Fig. 1a, but the
critical behavior at the tips is no longer that of $d+1$ dimensional
XY model. Unfortunately the tree-level RG equation (\ref{rg}) as well as
the more elaborate treatment that includes loop contributions of Ref. \onlinecite{sm}
cannot resolve which of the two scenarios is realized. This is because
the perturbative RG flow equation (\ref{rg}) is well controlled only at weak coupling or
small $\lambda_4$. While
it clearly establishes the instability of the $d+1$ dimensional XY fixed point ($z=1$)
when the
singular coupling $\lambda_4$ (due to presence of fermions) is present, one can not tell
where the RG flow brings the system to, unless a more stable fixed point (with only one
relevant perturbation) is found at weak coupling. On the other hand
it would be very interesting to use numerical methods like
quantum Monte Carlo to resolve which of the two scenario is
realized; in such studies particle-hole symmetry can be ensured by
fixing the boson number while varying $U/t_b$\cite{hebert}.

We now discuss case (ii) of Bose-Fermi mixture, in which the bosons
are molecules made of two fermions with opposite spins, with density
imbalance between them. Recently, much interest has focused on the
problem of pairing in such imbalanced fermionic
systems\cite{book,martin,randy}, with the hope of realizing novel
superfluid phases different from the BCS state\cite{yangreview}. In
the present paper we consider the relatively simple regime of strong
pairing (or BEC) regime, in which one can think that all the
minority (-) fermions are paired with majority (+) fermions to form
bosonic molecules, while the unpaired + fermions form a Fermi gas;
phase separation does not occur when the imbalance is small in this
regime\cite{sheehy}. As a result bosonic molecules coexist with
unpaired fermionic atoms. The simplest model that contains the basic
physics is the two-channel model on a lattice with one band for both
the bosons and fermions:
\begin{eqnarray}
H&=&H_{B}-t_f\sum_{<ij>,\sigma}(f^\dagger_{\sigma i}f_{\sigma
j}+h.c.) - \sum_{i\sigma} (\mu + \sigma h) n_{\sigma i}
\nonumber\\
& -& g\sum_{i}(f^\dagger_{+i}f^\dagger_{-i}b_i+h.c.).
\label{bosefermi2}
\end{eqnarray}
Here $\sigma=\pm 1$ is the (pseudo)spin index for the fermions,
$f_{\sigma i}$ is annihilation operator of the fermion on site $i$
with spin $\sigma$ and $n_{\sigma i}=f^\dagger_{\sigma i}f_{\sigma
i}$. The fermion species are at chemical potentials $\mu_{f \sigma}
= \mu + \sigma h$, while the boson is at chemical potential $\mu_b =
2\mu - \nu$. Here $h$ is the ``Zeeman field" which imbalances the
fermion densities, and $\nu$ is the ``detuning" across the Feshbach
resonance which scans between the BCS and BEC limits. As mentioned
earlier, we focus mainly in the regime of strong pairing or large
negative detuning, $|\nu|\gg g, \mu,t,U$. In this regime fermions
with opposite spins form closely bound molecules, and the system can
sustain considerable spin imbalance without phase
separation\cite{sheehy}. The unpaired majority ($+$) fermions form a
single Fermi surface\cite{sheehy,yang} with a volume satisfying a
generalized Luttinger theorem\cite{sy}.
It should be noted that while we have three types of particles, due
to the $g$ term in (\ref{bosefermi2}) there are only two conserved
charges: $N_\sigma=\sum_i(b_i^\dag b_i+f^\dagger_{\sigma i}f_{\sigma
i})$. The boson superfluid and insulating phases here are defined
with respect to the transport of conserved charge $N_-$. Physically
this reflects the fact that at low energy minority ($-$) fermions
exist in the form of constituents of bosonic molecules at
low-energy.

We can again trace out the fermions in Eq. (\ref{bosefermi2}), which
generates effective boson interaction terms of the form
(\ref{intfermi}) that are represented by a new set of diagrams of
Fig. 3. While the Hamiltonian (\ref{bosefermi2}) and the Feynman
diagrams Fig. 3 look very different from those of the previous case,
and in particular the superfluid order parameter ($b$) couples
directly to fermions here, the singularities generated by tracing
out fermions turn out to be {\em identical}.
Heuristically it can be seen from the following considerations. In
the Hamiltonian (\ref{bosefermi2}) the majority ($+$) fermions are
gappless but the minority ($-$) fermions remain {{\em fully gapped};
as a result the $-$ fermion Green's function has no pole at
low-energy and can be treated as a number ($-1/|\mu-h|$).
Diagramatically, one can thus shrink the $-$ fermion lines to a
point for the diagrams in Fig. 3, which reduces them to the same
diagrams as those in Fig. 2. Alternatively, one can also see this by
first integrating out the fully gapped $-$ fermions; this generates a
quartic (or density-density) interaction between bosons and gapless
+ fermions of the same form as in (\ref{bosefermi1}). They thus have
the same singularities, and we conclude that these two cases of
Bose-Fermi mixtures share the same phase diagram and critical
properties of SIT.

\begin{figure}
\begin{tabular}{cccccc}
\parbox[c]{2.3cm}{\includegraphics[scale=0.25]{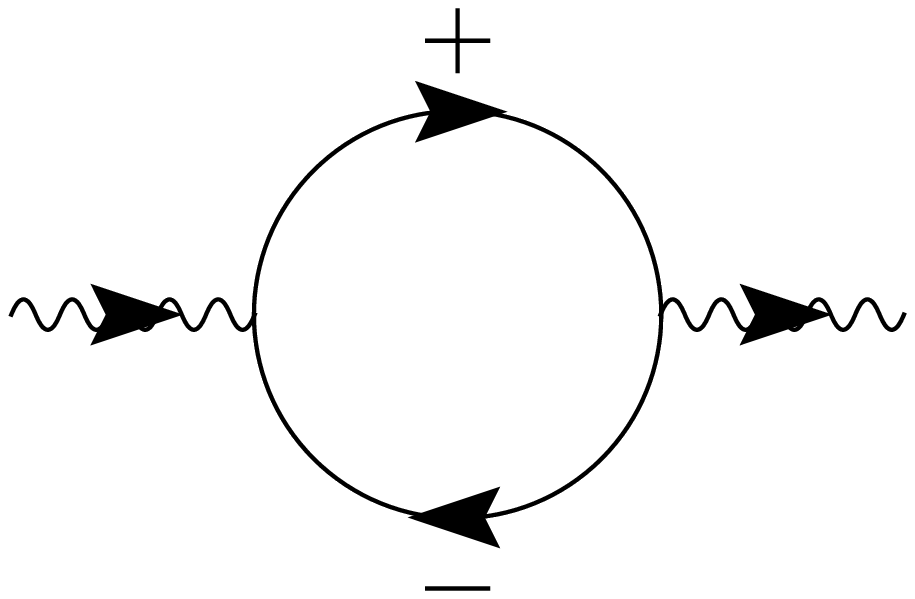}} & + &
\parbox[c]{1.7cm}{\includegraphics[scale=0.25]{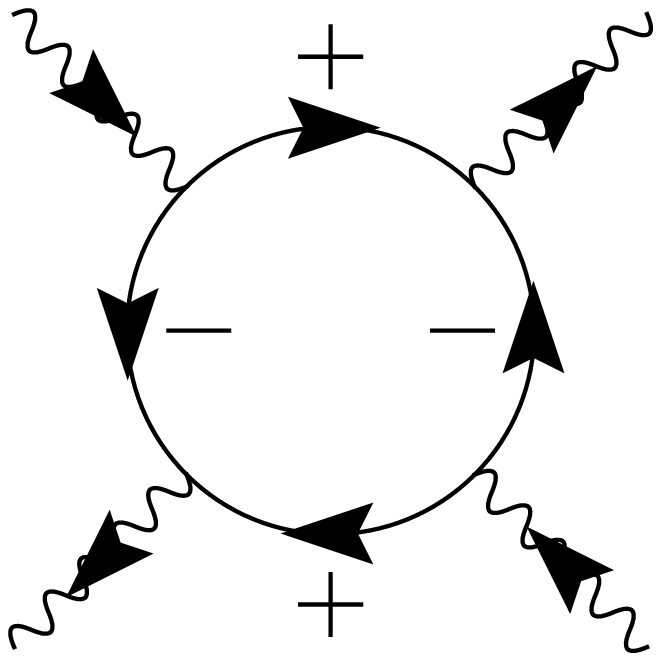}} & + &
\parbox[c]{2cm}{\includegraphics[scale=0.25]{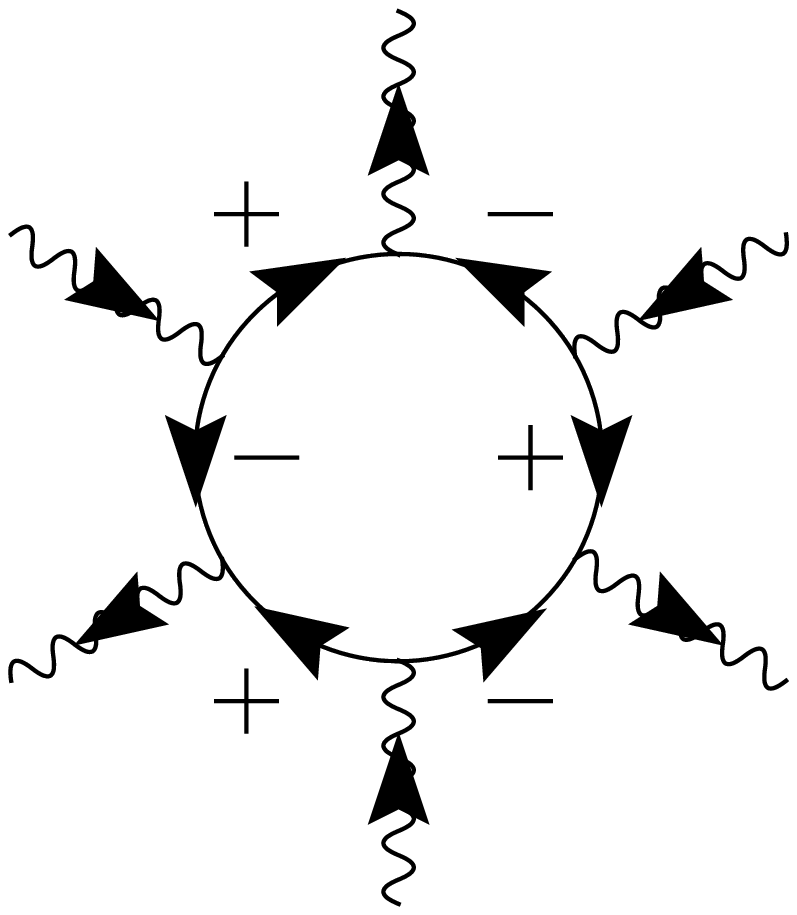}} & + $\cdots$
\end{tabular}
\caption{Same as Fig. 2, for case (ii) of Bose-Fermi mixture.}
\end{figure}

\section{fermion pairing near boson superfluid-insulator transition}

In this section we turn our discussion to the behavior of fermions, especially
near SIT. It should be clear from the discussion above that the
fermion physics is also similar for cases (i) and (ii), and we focus
on case (i) below. It has been known for a long time\cite{bardeen}
that boson density fluctuations mediate an attractive interaction
for case (i) of Bose-Fermi mixture, which at weak coupling takes the
form\cite{bijlsma,viverit}
\begin{equation}
V_{eff}({\bf q}, \omega)=V^2\chi({\bf q}, \omega),
\label{attraction}
\end{equation}
where $\chi({\bf q}, \omega)$ is the boson density susceptibility.
This attarctive interaction leads to fermion pairing, an effect that
has been studied in some detail for the case that the bosons are
deep in the superfluid phase\cite{bijlsma,viverit,efremov,pairingnote}; in this case the
effective interaction is dominated by single phonon exchange, and
takes a form very similar to the electron-phonon interaction in
metals. When the bosons are in the Mott insulator phase $\chi({\bf
q}, \omega)$ is zero at $T=0$ and small $\omega$, and takes an
activated behavior at low $T$ due to the Mott-Hubbard gap; as a
result the effective pairing interaction is extremely weak. In the
following we study the pairing physics near the particle-hole
asymmetric SIT with $z=2$, where the critical properties are known
and the Bose-Fermi interaction can be treated perturbatively due to
its irrelevance in the RG sense. We will comment on the case of
particle-hole symmetric SIT later on.

The critical behavior of particle-hole asymmetric SIT is that of the
dilute Bose gas transition\cite{subirnote}. For $d=2$ or $3$, the
theory is at or above its upper critical dimension, and a mean-field
(or Hartree-Fock) treatment of the boson interaction $U$ in the
calculation of $\chi({\bf q}, \omega)$ in Eq. ({\ref{attraction}) is
justifiable even near the transition. For simplicity we focus on
$d=3$ below to avoid complications associated with logarithmic
correction at the upper critical dimension $d=2$. For weak pairing
(which is assumed here), we may neglect the frequency dependence in
Eq. (\ref{attraction}) and reduce it to a static attractive
interaction:
\begin{equation}
V_{eff}({\bf q})=-{V^2\kappa\over 1+|{\bf q}|^2\xi^2},
\label{attraction1}
\end{equation}
where $\kappa={\partial n_B\over \partial \mu_B}$ is the boson
static compressibility, and $\xi$ is proportional to the boson
superfluid correlation length (with an $O(1)$ prefactor). This
attractive interaction leads to fermion superfluid transition
temperature in angular momentum channel $l$:
\begin{equation}
T^F_c= A_l E_F\exp\{-1/[D(E_F)\Gamma_l]\},
 \label{tcf}
\end{equation}
where $E_F$ is Fermi energy, $D(E_F)$ is fermion density of states
at $E_F$, $A_l$ is a constant of order 1 (e.g., $A_{l=0}\approx 0.61$), and
\begin{equation}
\Gamma_l={2l+1\over 2}
\int_0^{2\pi}{d\cos\theta}V_{eff}[2k_F\sin(\theta/2)]P_l(\cos\theta)
\end{equation}
is the pairing interaction in angular momentum channel $l$. Here
$k_F$ is the Fermi wave vector and $P_l$ is the Legendre polynomial.

On the superfluid side with $\Delta\mu > 0$ ($\Delta\mu$ is the
difference between $\mu_B$ and the phase boundary) and below the
boson superfluid transition temperature $T^B_c$, straightforward
calculation shows\cite{refereenote}
\begin{equation}
\label{k1}
\kappa\approx 1/U
\end{equation}
 which is mostly independent of
$\Delta\mu$, and
\begin{equation}
\label{xi1}
\xi\sim \sqrt{t_b/\Delta \mu},
\end{equation}
 which diverges
with mean-field exponent as $\Delta\mu\rightarrow 0$. The increasing
correlation length $\xi$ reduces the $s$-wave pairing strength
$\Gamma_0$. However it {\em enhances} the $p$-wave
pairing\cite{efremov} strength $\Gamma_1$ due to the stronger
momentum dependence of the pairing interaction, as long as
$\xi\lesssim 1/k_F$. Since for the single component fermion in our
model (\ref{bosefermi1}) $s$-wave pairing is not allowed and the
leading pairing instability is in the $p$-wave channel, we expect
fermion gap and $T^F_c$ to {\em increase} as one approaches SIT;
thus one expects $T^F_c$ to {\em peak} very close to the critical
point with $\Delta\mu =0$.

Since $T^B_c\rightarrow 0$ as $\Delta\mu \rightarrow 0$, $T^F_c$
will cross $T^B_c$ when one gets sufficiently close to the SIT. In
this case the pairing interaction is controlled by the finite $T$
properties of the $z=2$ QCP and becomes strongly $T$-dependent.
Again using mean-field approximation we obtain
\begin{equation}
\label{k2}
\kappa\sim
1/(U^2t_BT)^{1/4}
\end{equation}
and\cite{subirnote}
\begin{equation}
\label{xi2}
 \xi\sim
[t_b^5/(U^2T^3)]^{1/4}.
\end{equation}
In this case $T^F_c$ needs to be determined
self-consistently by solving Eq. (\ref{tcf}). Our qualitative
results on fermion pairing are summarized in Fig. 4.\cite{tcnote}

As discussed earlier, Bose-Fermi interaction is relevant at the
particle-hole symmetric QCP of SIT, and may flow to a strong
coupling fixed point. If this turns out to be the case then fermions
have a {\em universal} strong pairing interaction at this QCP, a
situation very similar to Feshbach resonance. We thus expect
$T^F_c\sim \tilde{E_F}$, where $\tilde{E_F}$ is the energy scale at
which the Bose-Fermi interaction has flown to the universal value.
We expect $\tilde{E_F}\sim E_F$ for $V\sim E_F$, as there is no
other energy scale for the fermions in this case. This could be an
alternative way to achieve ``high" $p$-wave $T^F_c$. In both the
speculative case here and the concrete case of particle-hole
asymmetric QCP with $z=2$ discussed above, we find $p$-wave $T^F_c$
peaks at the QCP due to pairing interaction mediated by quantum
critical fluctuations. This is similar to the quantum criticality
scenario of high $T_c$ superconductivity in cuprates, and provides a
concrete example of this scenario.

\begin{figure}
\includegraphics[width=0.45\textwidth]{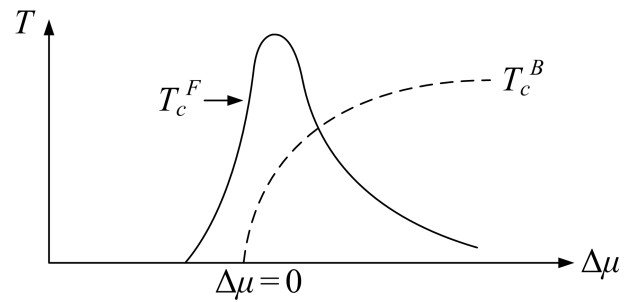}
\caption{Finite temperature phase diagram near the particle-hole
asymmetric boson superfluid-insulator transition, at $\Delta\mu=0$
and $T=0$. The solid line is $p$-wave fermion superfluid transition
temperature, while dashed line is boson superfluid transition
temperature.}
\end{figure}

\section{concluding remarks}

Bose-Fermi mixtures represent new quantum many-body systems
realized in trapped cold atomic gases, that do not have straightforward
correspondences in electronic condensed matter systems. In the latter case the constituent particles, namely electrons, are fermions. When the electrons form Cooper pairs, the system may also be described (at low energy) using boson only models like the Bose Hubbard model; in these cases typically the fermionic degrees of freedom are fully gapped and do not enter the low-energy effective theory.

In the present paper we have explored some new physics in Bose-Fermi mixtures, in which low-energy bosonic and fermionic excitations co-exist and interact with each other. More specifically, we have studied the effect of fermions on the critical properties of boson superfluid-insulator transition (SIT), and the pairing interaction between fermions
mediated by quantum critical fluctuations near SIT. In the latter case we predict the fermion pairing gap and superfluid transition temperature peak near boson SIT. It would be very interesting to use
quantum Monte Carlo or other numerical methods to test the
predictions of this paper, especially to resolve the nature of the
particle-hole symmetric boson SIT in the presence of fermions.

\acknowledgements
The author thanks Subir Sachdev and Fei Zhou for helpful
conversations, and Q. Cui for graphics assistance. Parts of this
work were performed at Kavli Institute for Theoretical Physics
(KITP, which was supported in part by the National Science
Foundation under Grant No. PHY05-51164), and Kavli Institute for
Theoretical Physics China (KITPC). This work was supported in by
National Science Foundation grants No. DMR-0225698 and No. DMR-0704133.


\begin{thebibliography}{1}
\bibitem{tcnote} While $T^F_c$ is determined by Eq. (\ref{tcf}), one needs knowledge of $\kappa$ and $\xi$ as input
to obtain its quantitaive value. Using the known critical properties of $z=2$ SIT we have obtained the exact
scaling behavior of $\kappa$ and $\xi$ in Eqs. (\ref{k1},\ref{xi1},\ref{k2},\ref{xi2}), but we cannot determine their non-universal prefactors which depend on
microscopic details. Thus our results on $T^F_c$ is only qualitative. However the conclusion that it peaks near SIT
is robust.
\end{thebibliography}
\end{document}